\documentclass{article}
\usepackage{amsmath,amssymb}
\usepackage{bm}
\usepackage{graphicx}
\usepackage{ascmac}
\newcommand{\bmX}{\bm{X}}
\newcommand{\bmx}{\bm{x}}

\begin{document}
%\begin{frontmatter}
\title{Application of autonomous pathfinding system to kinematics and dynamics problems by implementing network constraints} 
\author{Kei-Ichi Ueda \thanks{Faculty of Science, Academic Assembly, University of Toyama, Toyama 930-8555, Japan}}
%\fntext[label1]{Corresponding author: 3190 Gofuku, Toyama 930-8555,
%apan. Phone:+81\ 76\ 445\ 6561}
%\ead{kueda@sci.u-toyama.ac.jp}

%\address[label1]{Graduate School of
%Faculty of Science, Academic Assembly, University of Toyama, 
%Toyama 930-8555, Japan}
\date{}

\maketitle
\begin{abstract}
A neural network system in an animal brain contains many modules and generates 
adaptive behavior by integrating the outputs from the modules. 
The mathematical modeling of such large systems to elucidate the mechanism of rapidly finding solutions is vital to develop control 
methods for robotics and distributed computation algorithms. 
In this article, we present a network model to solve kinematics and dynamics problems for robot arm manipulation. 
This model represents the solution as an attractor in the phase space and 
also finds a new solution automatically when perturbations such as variations in the end position of the arm or obstacles occur. 
In the proposed model, the physical constraints, target position, and the existence of obstacles are represented by network connections. Therefore, the theoretical framework of the model remains almost the same when the number of constraints increases. 
In addition, as the model is regarded as a distributed system, it can be applied toward the development of parallel computation algorithms. 
\end{abstract}
%\begin{keyword}
%Kinematics problem; Dynamics problem; Pathfinding system; Dynamical system; Attractor
%\end{keyword}

%\end{frontmatter}

\section{Introduction}\label{sec:intro} 
Over the past few decades, trajectory control for robot arm manipulation has garnered considerable attention in robot engineering 
(Featherstone, R. \& Orin, D. (2000); Rodriguez, G., Jain, A., \& Kreutz-Delgado, K. (1992)). 
%\cite{Featherstone2000, Rodriguez1992}. 
A fundamental kinematics problem in such a control system is to determine a feasible position for the arm joint such that the end point approaches the target location. 
This issue is formulated in terms of boundary value problem (BVP) in two-dimensional space, which can be expressed as follows:
\begin{itemize}
\item[BVP:]
Find $\bmx_l \in Q_l$ ($l=2,3,\dotsc, L-1$) that satisfies the following conditions:
\begin{align*}
&|\bmx_l-\bmx_{l+1}| = d_l,\quad (l=1,\dotsc,L-1)\\
&\bmx_1 = (x_s, y_s) \in Q_1,\quad \bmx_L= (x_g, y_g) \in Q_L, 
\end{align*}
\noindent where $Q_l\subset \mathbb{R}^2$ is the feasible region for the joint $l$ and $L \geq 2$ is the segment number. 
\end{itemize}
The solution $\bmx_l$ corresponds to the position of $l^{th}$ joint in the two-dimensional space. 
As BVP has multiple solutions in general, we need to solve the singular equation of $\bm{x}_l$. 
In addition, if the obstacles are assumed to be placed in the domain or $Q_l\subsetneq \mathbb{R}^2$, 
additional constraints should be implemented in the equation. 
Numerical algorithms have been proposed for kinematics problems based on iterative methods 
%\cite{Aristidou2011,Unzueta2008} 
(Aristidou, A. \& Lasenby, J. (2011); Unzueta, L., Peinado, M., Boulic, R., \& Suescun, \'A. (2008))
and
neural network models 
%\cite{Tejomurtula1999,Koker2013,Toshani2014}. 
(Tejomurtula, S. \& Kak, S. (1999); K\"oKer, R. (2013); Toshani, H. \& Farrokhi, M. (2014)). 
For dynamics problems, we need to consider additional constraints to obtain a smooth arm motion. 
Several approaches have been proposed for solving dynamics problems such as
optimization methods 
%\cite{Wada1994,Poggio1990}, 
(Wada, Y., Koike, Y., Vatikiotis-Bateson, E., \& Kawato, M. (1994); Poggio, T. \&  Girosi, F. (1990)), 
self-organizing maps 
%\cite{Kuperstein1988,Walter1993}, 
(Kuperstein, M. (1988); Walter, J. A. \& Schulten, K. I. (1993)), 
neural network models 
%\cite{wada,Narendra1990,Glasius1995}, 
(Wada, Y. \& Kawato, M. (1993); Narendra, K. S. \& Parthasarathy, K. (1990); Glasius, R., Komoda, A., \& Gielen, S. C. A. M. (1995)), 
and reservoir computation 
%\cite{Polydoros2016}
(Polydoros, A. S., \& Nalpantidis, L. (2016)).

As the number of components in the system increases, the 
formulation of the algorithms becomes complicated in general. 
In this study, we propose a new framework for modeling kinematics and dynamics problems. Here, the solution of BVP is represented by a path in the network connecting nodes, which correspond to the boundary values. Further, the constraints in the arm length and the presence of obstacles are described by the addition and removal of the network links. Thus, the network construction procedure remains almost the same as the number of system's components and that of constraints increase. In fact, we need to attach or detach the network links according to the physical constraints and the position of obstacles.

Autonomy is an important concept that should be considered while constructing a robot system with adaptive behavior 
%\cite{Volpe2001}
(Volpe, R., Nesnas, I., Estlin, T., Mutz, D., Petras, R., \& Das, H. (2001)). 
When humans encounter unfamiliar environment, they autonomously develop strategies and execute new actions. 
There is an increasing demand for the development of such an autonomous system, whose control algorithm is based on just the variables of the system.
In addition, as the number of components in the system increases, 
the distributed processing is required to decrease the computation time. 
Therefore, the development of effective autonomous and distributed systems has received considerable attention in industries. 
If the system is formulated in terms of differential equations, the flexibility of the system against environmental variation can be regarded as the switching of attractor in the phase space. 
Thus, the elucidation of the mathematical mechanism for the robustness of attractor switching is vital to improve the performance of the system. Ueda et al. %\cite{Ueda2015} 
(Ueda, K. I., Yadome, M., \& and Nishiura, Y. (2015)) proposed a network model to show flexible attractor switching. 
This network model has been applied to pathfinding problems 
and shows the following properties: (1) The model can spontaneously find 
one of the possible paths connecting two target points. (2) It begins to find 
another path when perturbations such as removal of paths occur. 
Using the above properties and implementing network constraints, 
we construct a solver that can autonomously find a solution of BVP and finds another possible solution when the existing solution becomes impractical due to perturbations. 

To apply the pathfinding model to BVP, we formulate a discretized version of BVP, which is called DBVP. We define a two-dimensional 
lattice in $\widetilde{\Omega}$ 
and a set of the lattice points $\Gamma$ as follows: 
\begin{align*}
&\widetilde\Omega_l := \{\bmx \in \mathbb{R}^2\ |\ \bmx =(\xi_{i,j}, \eta_{i,j}),\  (i,j)\in \Gamma\},\quad l\in \{1,\dotsc, L\},\\
&\xi_{i,j} := x_{\min} + (i-1)\cdot (x_{\max}-x_{\min})/(J_x-1),\\
&\eta_{i,j} := y_{\min} + (j-1)\cdot (y_{\max}-y_{\min})/(J_y-1), \\
&\Gamma: = \{(i,j)\ |\ i=1,\dotsc, J_x, j=1,\dotsc, J_y\}. 
\end{align*}
Because we assume that the solutions are attained at the lattice points, 
we formulate the DBVP as follows: 
\begin{itemize}
\item[DBVP:]
Find $\bmX_l \in \widetilde{Q}_l\subset \widetilde{\Omega}_l$ ($l=2,3,\dotsc,L-1$) that satisfies the following conditions:
\begin{align}
\label{eqn:constd}
&|\bmX_l-\bmX_{l+1}|\in [d_l-\Delta d_l, d_l+\Delta d_l],\quad (l=1,\dotsc,L-1)\\
\label{eqn:dbvpbc}
\nonumber
&\bmX_1 = (x_s, y_s) \in \widetilde{Q}_1,\quad \bmX_L	= (x_g, y_g) \in \widetilde{Q}_L. 
\end{align}
\end{itemize}
In general, due to the discretization, we need to consider the margin $\Delta d_l$ as the constrain for $d_l$. The value of $\Delta d_l$ is determined by the geometrical constraint and can be reduced if $J_x$ and $J_y$ increase. 

Firstly, we apply the pathfinding model to DBVP. The boundary values for the base position $\bm{X}_1$ and the end position $\bm{X}_L$ are given 
as the start and target point in the network. The boundary condition and the physical constraint for the robot arm are described by the network. 
Secondly, we extend the DBVP model to the dynamics problem. 
As the network contains excitatory and inhibitory connections between the nodes and integration system does not exist, the model represents a distributed system. Therefore, our study is potentially useful for the development of the parallel computation algorithms to solve kinematics and dynamics problems.

\section{Pathfinding system}
We apply the model proposed in 
%\cite{Ueda2015}
Ueda, K. I., Yadome, M., \& and Nishiura, Y. (2015), which can find one of the 
possible paths connecting the start and target points in hierarchical network consisting of nodes and directional excitatory and inhibitory links. 
The node dynamics is described by differential equations. The 
solution path is described by a stationary state of the model. 

\subsection{Network construction} 
\begin{figure*}[htbp]
 \begin{center}
 \includegraphics[keepaspectratio=true, width=0.90\textwidth]{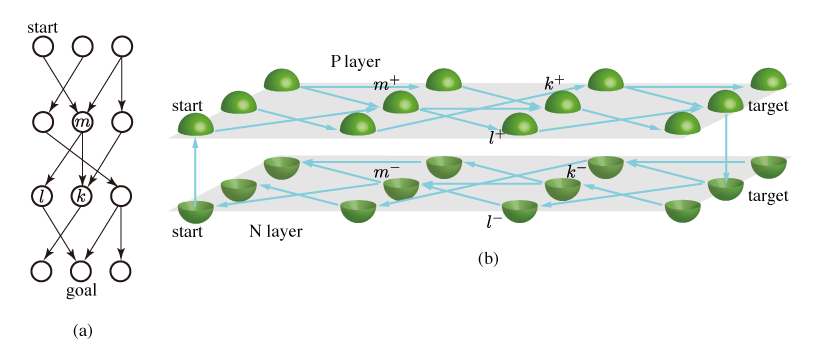}
\end{center}
 \caption{(a) An example of hierarchical network. 
 The start and target nodes are placed at the top and bottom layer of the hierarchy. 
 (b) A network model for the network in (a). Arrows indicate 
 the excitatory links. The direction of the excitatory link in the N layer is opposite to 
 that in the P layer. Additional excitatory links are added at the start and target nodes. 
}
\label{fig:concept}
\end{figure*}

The network-construction procedure of the pathfinding system based on a hierarchical network is shown in Fig.~ \ref{fig:concept}. 
We assume that the start and target points of the network  are at the top and bottom layer, respectively. 
Excitatory and inhibitory links are attached according to the following rules: 
\begin{itemize}
\item[(P1)] 
The nodes corresponding to the point $k$ 
in the network (Fig.~ \ref{fig:concept}(a)) are placed at the P and N layers (Fig.~ \ref{fig:concept}(b)) . These nodes are 
called {\it node $k^+$} and {\it node $k^-$}, respectively. 
\item[(P2)] The excitatory links directed from node $m^+$ to $k^+$ and from $m^-$ to $k^-$ are attached (Fig.~ \ref{fig:concept}(b))
if a connection exists between point $m$ and $k$ (Fig.~ \ref{fig:concept}(a)). 
The existence of excitatory interaction directed from nodes $m^\pm$ to $k^\pm$ is represented as $a_{m^{\pm},k^{\pm}}$, where $a_{m^{\pm},k^{\pm}}=1$ and $a_{m^{\pm},k^{\pm}}=0$ indicate 
the presence and absence of such interactions, respectively.

\item[(P3)] There are inhibitory links from nodes $m^+$, 
	$l^+$, and $l^-$ to node $k^+$ and from 
	nodes $m^+$, $k^+$, and $k^-$ to node $l^+$
	if there are excitatory links from 
	node $m^+$ to both the nodes $k^+$ and $l^+$. 
	Similarly, there are inhibitory links from nodes $m^-$, 
	$l^-$ and $l^+$ to node $k^-$ and from 
	nodes $m^-$, $k^-$ and $k^+$ to node $l^-$ 
	if there are excitatory links from 
	node $m^-$ to the nodes $k^-$ and $l^-$. 
\end{itemize}
According to the above procedure, the activated state, which is defined as ON state, propagates from top to bottom in the P layer and from bottom to the top in the N layer.

We add excitatory links between P and N layers at the start and target nodes to 
form a loop. Thus, the solution path connecting the start and target nodes is represented by the 
nodes with ON state forming a loop network architecture. 
We use two different descriptions for the target point, which are discussed in Sec. \ref{sec:dynamics}. 
\begin{itemize}
\item[(P4)] 
 The network has an excitatory link from node
	 $k_s^-$ to node $k_s^+$ and from node $k_g^+$ to node $k_g^-$, where 
	 $k_s^+$ ($k_s^-$) and $k_g^+$ ($k_g^-$) indicate the nodes at the start and target positions in the P (N) layer, respectively. The boundary condition for the start position is expressed in terms of the link connection
\begin{equation*}
\hat{a}_{k^-, k^+} =
\begin{cases}
1\quad &\text{if $k^\pm = k^\pm_s$}\\
0\quad &\text{otherwise}
\end{cases}
\end{equation*}
We employ two types of boundary conditions for the target position, which are expressed as follows: 
\begin{align}
\label{eqn:bcmodel1}
\hat{a}_{k^+, k^-} &=
\begin{cases}
1\quad &\text{if $k^\pm = k^\pm_g$}\\
0\quad &\text{otherwise}
\end{cases}
,\\
\label{eqn:bcmodel2}
\check{a}_{k} &=
\begin{cases}
1\quad &\text{if $k = k^-_g$}\\
0\quad &\text{otherwise}
\end{cases}
.
\end{align}
\end{itemize}
\subsection{Model formulation}
\begin{figure*}[htbp]
 \begin{center}
 \includegraphics[keepaspectratio=true, width=0.90\textwidth]{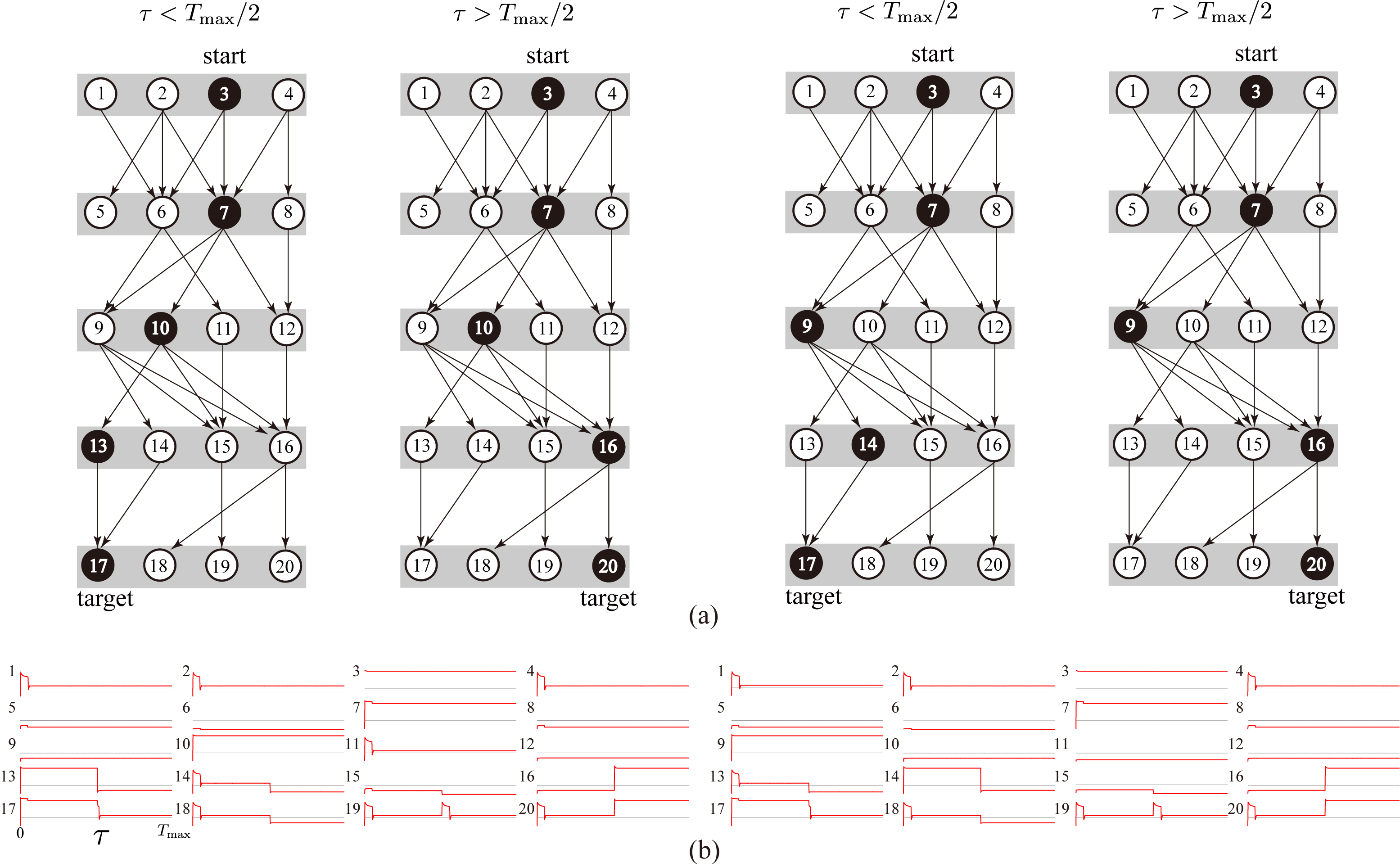}
\end{center}
 \caption{Numerical results obtained by Model I (Left) and II (Right). (a) An example of network structure, where only the excitatory links in the P layer are shown. 
 The target position is considered at node 17 for $\tau<T_{\max}$ and at 20 for $\tau>T_{\max}$. 
 Circles represent the node state when the solution converges to stationary state for $\tau<T_{\max}$ and $\tau>T_{\max}$. 
Black and white circles indicate the ON and OFF states, respectively. 
(b) Numerical solution of equation \eqref{eqn:model} for the network. Only the time sequences of the nodes in P layer are shown. 
The horizontal and vertical directions indicate $u_{k^+}$ and $\tau$, respectively. 
 It is clear that the system successfully finds a possible solution when the target position is varied. 
}
\label{fig:path}
\end{figure*}

The pathfinding models with boundary conditions given by equations \eqref{eqn:bcmodel1} and \eqref{eqn:bcmodel2} are referred to as Model I and Model II, respectively. 
According to the rules (P1) - (P4), Model I is described by 
\begin{equation} \label{eqn:model}
\begin{aligned}
  &\dot{u}_{k^\pm} =  f(u_{k^\pm},v_{k^\pm}) + \mu_1
   H_0\left(\sum_{m^{\pm}\in\Lambda^\pm} a_{m^\pm,k^\pm}
 H_0(u_{m^\pm}-\theta_1) 
 \right)\\
 &\quad+ \mu_1  \hat{a}_{k^\mp, k^\pm}H_0(u_{m^\pm}-\theta_1)\\
 &\quad - \mu_2  H_0\left(\sum_{m^\pm\in\Lambda^\pm}\sum_{l\in\Lambda}
   a_{m^\pm,l^\pm}a_{m^\pm,k^\pm}
H_0(u_{m^\pm}-\theta_1) H_0(u_{l^\pm}+u_{l^\mp}-\theta_2)\right)\\
 &\quad +A\sigma_{k^\pm}(t),
   \vspace{0.2cm}\\
  &\dot{v}_{k^\pm} = g(u_{k^\pm},v_{k^\pm}). \hspace{0.7cm}
\end{aligned}
\end{equation}%
Similarly, Model II is described by
\begin{equation} \label{eqn:model2}
\begin{aligned}
  &\dot{u}_{k^+} =  f(u_{k^+},v_{k^+}) + \mu_1
   H_0\left(\sum_{m^{+}\in\Lambda^\pm} a_{m^+,k^+}
 H_0(u_{m^+}-\theta_1) 
 \right)\\
 &\quad+ \mu_1  \hat{a}_{k^-, k^+}H_0(u_{m^\pm}-\theta_1)\\
 &\quad - \mu_2  H_0\left(\sum_{m^+\in\Lambda^+}\sum_{l\in\Lambda}
   a_{m^+,l^+}a_{m^+,k^+}
H_0(u_{m^+}-\theta_1) H_0(u_{l^+}+u_{l^-}-\theta_2)\right)\\
 &\quad +A\sigma_{k^+}(t),
   \vspace{0.2cm}\\
  &\dot{v}_{k^+} = g(u_{k^+},v_{k^+}), \\
  &\dot{u}_{k^-} =  f(u_{k^-},v_{k^-}) + \mu_1
   H_0\left(\sum_{m^{-}\in\Lambda^-} a_{m^-,k^-}
 H_0(u_{m^-}-\theta_1) 
 \right)\\
 &\quad + \mu_1  \check{a}_{k^-} \\
 &\quad - \mu_2  H_0\left(\sum_{m^-\in\Lambda^-}\sum_{l\in\Lambda}
   a_{m^-,l^-}a_{m^-,k^-}
H_0(u_{m^-}-\theta_1) H_0(u_{l^-}+u_{l^+}-\theta_2)\right)\\
 &\quad +A\sigma_{k^-}(t),
   \vspace{0.2cm}\\
  &\dot{v}_{k^-} = g(u_{k^-},v_{k^-}), \hspace{0.7cm}
\end{aligned}
\end{equation}%
where $k^\pm = 1^\pm,2^\pm,\dots,K^\pm$, $\Lambda^\pm:=\{1^\pm, 2^\pm,\dotsc, K^\pm\}$,
$\Lambda:=\{1, 2,\dotsc, K\}$, and
$\tau\in [0, T_{\max}]$ is
dimensionless time. The dot above $u$ and $v$ indicates their derivative with respect to $\tau$, and $\mu_i$ and $\theta_i$ ($i=1,2$) 
are positive constants. The functions $f$ and $g$ are described by the sigmoid FitzHugh--Nagumo
equation 
%\cite{rots}
(Rotstein, H. G., Kopell, N., Zhabotinsky, A. M., \&. Epstein, I. R. (2003))
: $f(u, v)=-a u^3 + b u^2 - c u + d - v$,
$g(u,v)=\epsilon [p\tanh\left((u-q)/r\right) + s - v]$.
We set $(a,b,c,d,p,q,r,s,\epsilon)=(1.92,4.32,1.8,0.1,0.72,0.3,0.2,0.261,0.03)$. 
The distance determines the time duration of activated state induced by the post-inhibitory 
rebound (PIR) behavior, which is explained in \ref{sec:append2}. 
The second and third terms on the right-hand side of the
equations \eqref{eqn:model} and \eqref{eqn:model2} correspond to the excitatory interactions, 
and the fourth term corresponds to the inhibitory interactions; 
$\sigma_{k^\pm}$ represents Gaussian noise with zero mean and unit variance, and $A$ is the noise amplitude. 
We assume that the interaction function $H_0$ is the Heaviside step
function and has a threshold $\theta$ such that the connectivity of the
link switches dynamically: $H_0(u-\theta)=1$ for
$u>\theta$ and $H_0(u-\theta)=0$ otherwise. 
The third term on the right-hand side of $u_{k^\pm}$-equation ensures that the system finds a single path from the multiple feasible solutions with the same 
route at the P and N layer. 
Based on this formulation, the node $k^+$ receives an inhibitory input 
when $u_{m^+}>\theta_1$ and $u_{l^+}+u_{l^-}>\theta_2$. 
For each node, we define ON, OFF1, and OFF2 states depending on $u_k$. 
The definition of the node states and the process for finding the solutions are described in \ref{sec:append1} and \ref{sec:append2}. 
Here, we call the OFF1 and OFF2 states as OFF state. 
To recapitulate, the system described by equations \eqref{eqn:model} and \eqref{eqn:model2} has the following properties:
\begin{itemize}
\item The system robustly finds one of the possible paths connecting the start and target nodes if the solution exists. The associated nodes in P and N layers acquire ON state. 
\item All the nodes acquire OFF state if no solution exists. This implies that the system can terminate the search process if no possible solution exists. 
\item The system automatically starts the search process when the existing path is damaged, and also terminates the search process when it finds a new possible path. 
\end{itemize}
The representative time sequences of $u_{k^+}$ for Model I and Model II are shown in Fig.~ \ref{fig:path}. 
It is evident that the models successfully find solutions and exhibit flexible attractor switching when the target position is changed. 

\begin{figure*}[htbp]
 \begin{center}
 \includegraphics[keepaspectratio=true, width=0.60\textwidth]{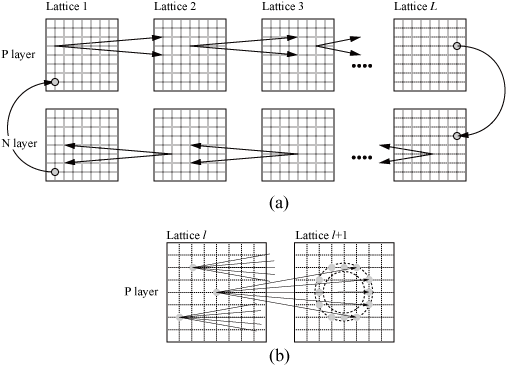}
\end{center}
 \caption{Schematic of excitatory link network. (a) According to ($\text{P2}^\prime$), 
excitatory links are connected from $l$ to $l+1$ in the P layer and 
from $l+1$ to $l$ in N layer. 
To form a loop structure of excitatory links, the links representing the start and end positions are installed from N to P layer and from P to N layer, respectively. 
(b) Excitatory links are attached if the corresponding nodes satisfy the constraint in equation \eqref{eqn:constd}. 
}
\label{fig:link}
\end{figure*}

\section{DBVP solver}
The solution of DBVP is represented by a path connecting the start and target nodes, which correspond to the base and end points, respectively. 
The nodes are placed on a two-dimensional square lattice and the solution of DBVP, i.e., $\bm{X}_l$ is represented by the position of nodes in ON state. 
The physical constraint and the existence of the obstacles are described by link connection and disconnection. 

The network consists of $L$ pairs of lattices and the 
$l^{th}$ pair is used to represent the position of $l^{th}$ joint. 
The lattice size is $J_x\times J_y$, where $J_x$ and $J_y$ 
represent the grid size of the $x$- and $y$-coordinates, respectively. 
Thus, the resolution of the approximation method is improved as $J_x$ and $J_y$ increase. 
Nodes are located at every lattice point. Therefore, the total number of nodes is $2 J_x J_yL$. 
The minimum and maximum values of the $x$-coordinate ($y$-coordinate)
are $x_{\min}$ ($y_{\min}$) and $x_{\max}$ ($y_{\max}$), respectively.
For notational convenience, the serial number of each node in given in each layer. 
To distinguish whether the node belongs to P or N layer, 
the $k^{th}$ node in P and N layer is called $k^+$ and $k^-$ node, respectively. 
We define a set of the serial number of nodes located in the $l^{th}$ lattice in P and N layer as $\Lambda^+(l)$ and 
$\Lambda^-(l)$, respectively, i.e.,
\begin{equation*}
\Lambda^\pm(l) = \left\{[1 + (l-1)\cdot J_xJ_y]^\pm, [2 + (l-1)\cdot J_xJ_y]^\pm, \dotsc, [J_xJ_y + (l-1)\cdot J_xJ_y]^\pm\right\}.
\end{equation*}
The $x$- and $y$-value of the node $k^\pm(i,j,l)\in \Lambda^\pm(l)$ are defined by 
\begin{align*}
&\bm{\xi}_{k^\pm} = \bm{\xi}_{k^\pm(i, j, l)} = (\xi_{k^\pm (i,j,l)}, \eta_{k^\pm (i,j,l)}),\\
&\xi_{k^\pm(i,j,l)} := x_{\min} + (i-1)\cdot (x_{\max}-x_{\min})/(J_x-1),\\
&\eta_{k^\pm(i,j,l)} := y_{\min} + (j-1)\cdot (y_{\max}-y_{\min})/(J_y-1).
\end{align*}
Model I is applied to DBVP by modifying (P2) and (P4) as follows: 
\begin{itemize}
\item[($\text{P2}^\prime$)]
The links in P layer are attached if the two corresponding nodes satisfy 
the physical constraint in equation \eqref{eqn:constd} (Fig.~ \ref{fig:link}). This implies that
\begin{equation*}
\begin{aligned}
&a_{m^+,k^+} = 
\left\{
\begin{aligned}
&1\quad \text{if}\quad |\bm{\xi}_{m^+} - \bm{\xi}_{k^+}| \in [d_l - \Delta d_l, d_l + \Delta d_l],\ m^+\in \Lambda^+(l),\\
&\qquad\quad k^+\in \Lambda^+(l+1) ,\ l\in [1,\dotsc, L-1]\\
&0\quad \text{otherwise}
\end{aligned}
\right.
\end{aligned}
\end{equation*}
Excitatory links in N layer, i.e., $a_{k^-,m^-}$ ($k^-\in \Lambda^-(l+1),\ m^-\in \Lambda^-(l)$) 
are determined according to (P2). 
\item[($\text{P4}^\prime$)]
The node numbers for the start (target) node in P and N layer are denoted as $k_s^+$ ($k_g^+$) and $k_s^-$ ($k_g^-$), respectively. 
According to (P4), the excitatory links are attached from the target node in P layer to the target node in N layer and 
from the start node in the N layer to the target node in P layer, i.e.,
\begin{equation*}
\hat{a}_{k^-, k^+} =
\begin{cases}
1\quad &\text{if $k^\pm = k^\pm_s$}\\
0\quad &\text{otherwise}
\end{cases}
,\quad 
\hat{a}_{k^+, k^-} =
\begin{cases}
1\quad &\text{if $k^\pm = k^\pm_g$}\\
0\quad &\text{otherwise},
\end{cases}
\end{equation*}
where $k_s^\pm \in \Lambda^\pm(1)$ and $k_g^\pm \in \Lambda^\pm(L)$ are the boundary conditions. 
Due to the inhibitory interaction, only a single pair of nodes acquire ON state for every
$l\in \{1,\dotsc, L \}$ when the system finds a solution. 
 \end{itemize}
Due to these assumptions, the solutions of the model necessarily satisfy the physical constraints and boundary conditions. 

\begin{figure}[htbp]
 \begin{center}
 \includegraphics[keepaspectratio=true, width=0.9\textwidth]{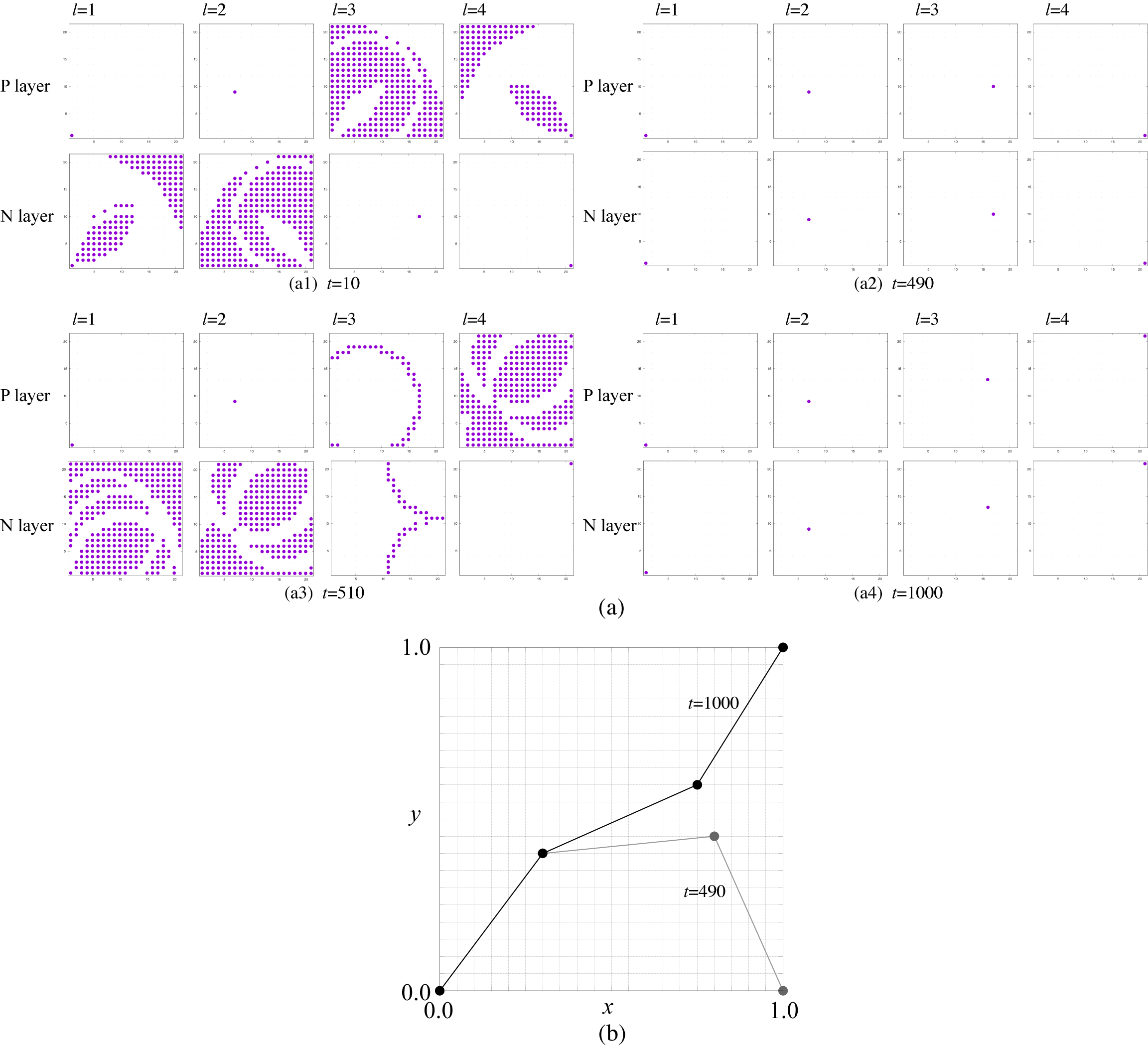}
\end{center}
 \caption{(a) Time sequence of the node state in P and N layers. The circles indicate ON state. (b) The positions of the joints are displayed in the lattice (black: $\tau= 490< T_{\max}/2$, gray: $\tau= T_{\max}=1000$). Here, $L=4$ and $l = 1.5/(L-1) = 0.5$.
 } 
\label{fig:handmove}
\end{figure}

\begin{figure}[htbp]
 \begin{center}
 \includegraphics[keepaspectratio=true, width=0.8\textwidth]{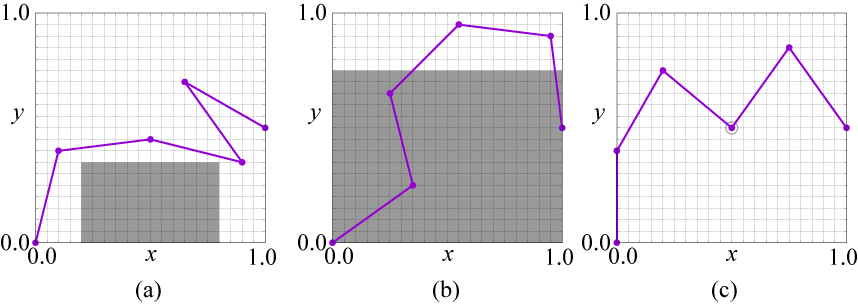}
\end{center}
 \caption{The position of $l^{th}$ joint is displayed in the lattice. (a) Case 1. 
 (b) Case 2. (c) Case 3. The 
 gray regions in (a) and (b) indicate the forbidden region $\widetilde{Q}^c$ for 
 all the joints and for joints $4$ and $5$, respectively. 
 The feasible region for joint $4$ is indicated by circles.}
\label{fig:solution}
\end{figure}

\section{Numerical results}
We consider the following cases as perturbations: (1) variation of boundary value $\bm{X}_L$ during computation, and (2) the existence of obstacles. In this section, for simplicity, we consider that $x_{\min}=y_{\min} = 0$, $x_{\max}=y_{\max} = 1$, and $J_x=J_y=21$. The Euler--Maruyama method is used for time integration, where 
the time grid is set as $\Delta \tau = 0.01$. 
For obtaining the approximate solution, 
$\Delta d_l$ should be determined so that the union of the circles covers the entire region of the solution space. We consider $\Delta d_l= \sqrt{2} \Delta x \ (=\sqrt{2}\Delta y)$ so that 
the model represented by equation \eqref{eqn:model} can robustly find a solution, where $\Delta x = \Delta y = 1/J_x = 0.1$. 
The parameters are set as $\theta_1=1.1$, $\theta_2=3.63$, $\mu_1=1.6$, $\mu_2=9.0$, and $A=1.0\times 10^{-4}$. 
As an initial state, the node at the start position in the P layer is considered to be in ON state and the other nodes are in OFF state for all experiments. 

\subsection{Adaptability of the solution-finding process} 
%We assess the case of $L=6$ and $l=0.2$. 
To confirm that the system can flexibly find a new solution, the 
boundary value $\bm{X}_L$ is changed during the computation. 
The boundary values $\bm{X}_1$ and $\bm{X}_L$ are given by 
\begin{equation*}
\bm{X}_1 = (0,0),\quad 
\bm{X}_L = 
\begin{cases}
(1, 0) & \text{for}\ \tau<T_{\max}/2,\\
(1,1) & \text{otherwise},
\end{cases}
\end{equation*}
where $T_{\max}= 1000$. 
Initially, ON state propagates from layer 1 to $L$ in the P layer and then from layer $L$ to $1$ in the N layer (Fig.~ \ref{fig:handmove} (a1)). 
The system successfully finds one of the possible solutions before $\tau=T_{\max}/2$, i.e., only one node is in 
ON state at every layer (Fig.~ \ref{fig:handmove} (a2)). 
The transient dynamics is observed just after the position of $\bm{X}_L$ is changed (Fig.~ \ref{fig:handmove} (a3)). The system successfully finds a new solution before $\tau=T_{\max}$ (Fig.~ \ref{fig:handmove} (a4)). 

\subsection{Obstacle avoidance}
Here, we consider the case in which obstacles are placed in the system.
The existence of the obstacles is represented by the removal of 
excitatory links directed to the nodes located at the positions of obstacles. 
We assume that the links emanating from the nodes in the forbidden 
region are removed in the P layer, and the links directed to the nodes in the forbidden region are removed in the N layer. This implies that
\begin{equation*}
a_{k^+,m^+} = a_{m^-,k^-} = 0,\quad \text{if}\quad (\xi_{k^+}, \eta_{k^+})\in \widetilde{Q}_l^c \quad (l=1,\cdots, L),
\end{equation*}
where $\widetilde{Q}_l^c:=\widetilde{\Omega} \backslash \widetilde{Q}_l$. 

Other connections are determined according to ($\text{P2}^\prime$) and ($\text{P3}$). 
We examine numerical results for the following three cases. 
For all the cases, we consider that $L=6$, $l = 0.4$, $T_{\max}=500$, $\bm{X}_1=(0,0)$, and $\bm{X}_L=(1.0,0.5)$. 
\subsubsection*{Case 1}
The forbidden region is given by 
\begin{equation*}
\widetilde{Q}_l^c = \{(x, y)\ |\ x\in [0.2, 0.8], y\in[0, 0.5]\},\quad l\in \{1,\dotsc, L\} 
\end{equation*}

\subsubsection*{Case 2}
The forbidden region for the joints $6$ and $7$ is given by 
\begin{equation*}
\widetilde{Q}_l^c = 
\begin{cases}
& \{(x,y)\ |\ x\in [0,1], y\in[0, 0.8]\},\quad \text{if}\quad l = 6,7\\
& \emptyset,\quad \text{otherwise}
\end{cases}
.
\end{equation*}

\subsubsection*{Case 3}
This case corresponds to the combination of the constrains in the above cases. Some joints are restricted to a specific position or to a specific region. For example, we consider the following constraint:
\begin{equation*}
\widetilde{Q}_l^c = 
\begin{cases}
& \widetilde{\Omega}\backslash\{(x,y)\ |\ x=0.5, y=0.5\},\quad \text{if}\quad l = 4\\
& \emptyset,\quad \text{otherwise}
 \end{cases}.
\end{equation*}
This implies that $\widetilde{Q}_4=\{(x,y)\ |\ x=0.5, y=0.5\}$. 
It may be noted that $(0.5,0.5)\in \widetilde{Q}_l$. 
Figure \ref{fig:solution} shows the numerical solution for these three cases. 
It is clear that the model successfully finds one of the possible solutions satisfying the constraints. 

\subsection{Avoidance of obstacle motion} 
\begin{figure}[htbp]
 \begin{center}
 \includegraphics[keepaspectratio=true, width=0.7\textwidth]{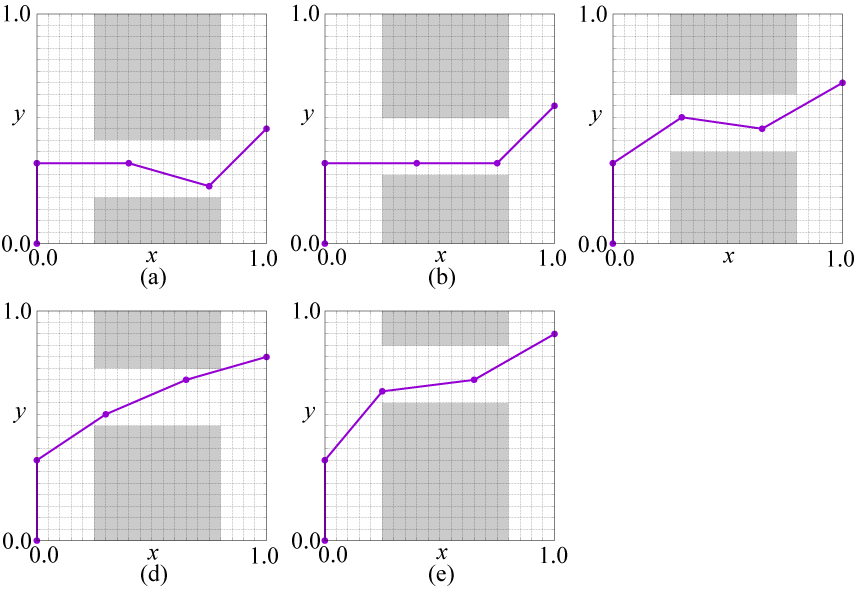}
\end{center}
 \caption{Transition of the solution of DBVP when the obstacles move according to equation \eqref{eqn:moving}. 
 The gray regions indicate the forbidden region. 
 Transient dynamics is initiated when $\tau=10000\times (m+1)$ ($m=0, 1,\dotsc$). 
 The model finds a new solution before $t$ reaches $10000\times m + 5000$ for every $m$. Here, $L=4$ and $l=1.5/(L-1)=0.5$. Solution for (a) $m=0$, (b) $m=2$, (c) $m=4$, (d) $m=6$, and (e) $m=8$. %(f) All the solutions for $m=1,2,\cdots, 10$. 
 }
\label{fig:obsmove}
\end{figure}

The variation in the position of obstacle is represented by the attachment and removal of links. As established in earlier studies 
%\cite{Ueda2015}
(Ueda, K. I., Yadome, M., \& and Nishiura, Y. (2015)), 
the 
proposed model can autonomously find a new solution when the network structure varies during the computation. 
Owing to this property, the system autonomously begins to find new solution 
when an obstacle destroys the existing solution. We assume that the obstacle motion is described by 
\begin{equation}\label{eqn:moving}
\begin{aligned}
&\widetilde{Q}_l^c(\tau) = \{(x,y)\in\mathbb{R}^2\ |\ x\in [0.4,0.6]\cap \widetilde{\Omega}_l, 
y\in ([0, 0.2+\varphi_\tau]\cup[0.45+\varphi_\tau, L_y])\cap \widetilde{\Omega}_l\},\\
&\bm{X}_1 = (0, 0) \in \widetilde{\Omega}_1,\ \bm{X}_L = (1.0, 0.5+0.05\varphi_\tau)\in \widetilde{\Omega}_L,\\
&\varphi_\tau = 0.05\times \left\lfloor \frac{\tau}{10000} \right\rfloor,
\end{aligned}
\end{equation}
where $\lfloor x \rfloor=\max\{n\in \mathbb{Z}\ |\ n\leq x\}$. 
Figure \ref{fig:obsmove} proves that the model autonomously starts to find another solution when the existing solution enters the forbidden region, 
and every joint corresponding to the new solution enters the feasible region.

\subsection{Number of steps required to find the solutions}\label{sec:time}
\begin{figure}[htbp]
 \begin{center}
 \includegraphics[keepaspectratio=true, width=0.4\textwidth]{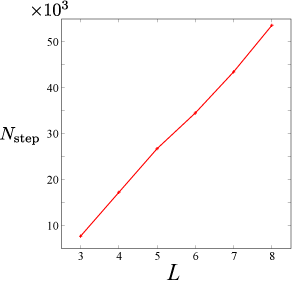}
\end{center}
 \caption{Average number of steps required to find a solution across 20 trials as a function of $L$.}
\label{fig:ave}
\end{figure}

Here, we measure the rate of increase in the number of steps when $L$ is increased. The number of steps is defined as $N_{\text{step}}$ when the solution converges to a stationary state or when a single pair of the nodes acquires ON state for every pair in the layers. 
This implies that if the solution is found in time $T'$, then $N_{\text{step}} = T'/\Delta \tau$. 
The positions $(x_1,y_1)$ and $(x_L,y_L)$ are fixed at $(0,0)$ [$(i,j) = (1,1)$] and 
$(1,0.5)$ [$(i,j) = (21,11)$], respectively. 
We measured the number of steps across 20 trials by using random seeds and 
calculated their average. 
It is noted that the number of steps does not indicate the actual computation time. 
From the search process of the model shown in \ref{sec:append2}, it is expected that the total number of steps is essentially determined by the number of steps during one round trip between the start and target points. This implies that the number of steps should be linearly proportional to $L$, which is confirmed in Fig.~ \ref{fig:ave}.

\section{Application of the pathfinding model to dynamics problem}\label{sec:dynamics}
We extend the model for the motion problem to solve the orbit problem. 
It is not guaranteed that the system represented by equation \eqref{eqn:model} exhibits a smooth motion when the target 
position is given because only physical and boundary conditions are employed as constraints. 
Therefore, to apply our algorithm for generating smooth trajectories of robot arm, additional constraint with respect to the continuity of feasible regions is required, which is called {\it time constraint}. 

\begin{figure}[htbp]
 \begin{center}
 \includegraphics[keepaspectratio=true, width=0.9\textwidth]{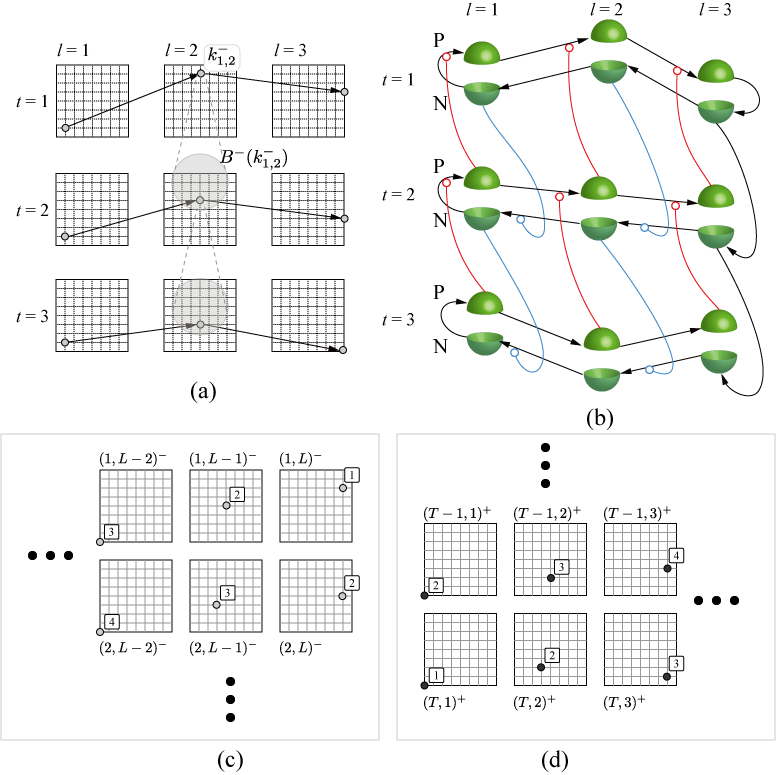}
\end{center}
 \caption{Schematic of the network model. (a) The arrows indicate the excitatory links in the P layers. 
 Gray circles indicate the time constraint $B^-(k_{1,2})$. (b) Three-dimensional network structure 
 for the excitatory links of (a). 
(c) The gray circles indicate that only the nodes in the N layer are in ON state. 
%The ON state propagates from $(1, L)^-$ to $(1, 1)^-$ and from $(1, L)^-$ to $(1, 1)^-$. 
(Step 1) Due to the initial data, the node at the target point in $(1, L)^-$ layer 
acquires ON state. (Step 2) The node at $(1, L-1)^-$ and $(2, L)^-$ layer acquires ON state. 
(Step 3) As both $(1, L-1)^-$ and $(2, L)^-$ layer are in the ON state, 
the node at $(2, L-1)^-$ also acquires ON state. In addition, the node at $(1,L-2)^-$ acquires ON state. 
(Step 4) As both $(1, L-2)^-$ and $(2, L-1)^-$ layer are in ON state, 
the node at $(2, L-2)^-$ acquires ON state. 
(d) The black circles indicate that nodes in both P and N layers are in ON state. 
(Step 1) The node at the start point of $(t,1)^+$ layer ($1\leq t\leq T-1$) can be in ON state 
when both the nodes at the start in $(t, 1)^-$ and $(t+1, 1)^+$ acquire ON state, 
but $(T, 1)^+$ acquires ON state when $(T, 1)^-$ is in ON state. 
Thus, the node at the start point $(T,1)^+$ acquires ON state faster than the other nodes in P layer. 
(Step 2) Similar to the case of N layer, a possible pair in $(T, 2)^+$ and 
$(T-1, 1)^+$ acquires ON state. (Step 3) Subsequently, a possible pair in $(T, 3)^+$ and $(T-1, 2)^+$ acquires ON state. 
(Step 4) As both $(T, 3)^+$ and $(T-1, 2)^+$ layer are in ON state, 
the node at $(T-1, 3)^+$ acquires ON state. 
}
\label{fig:kaiso}
\end{figure}

We consider the following DBVP with time constraint (DBVPT). 
\begin{itemize}
\item[DBVPT:]
Find $\bmX_{t, l} \in \widetilde{Q}_l$ ($t=1,\dotsc, T; l=1,\dotsc,L$) that satisfies the following conditions:
\begin{align}
\label{eqn:physt}
&|\bmX_{t, l}-\bmX_{t, l+1}|\in [d_l-\Delta d_l, d_l+\Delta d_l],\quad (t=1,\dotsc, T;\ l=1,\dotsc,L-1),\\
\label{eqn:timet}
&|\bmX_{t, l}-\bmX_{t+1, l}|\in [0, \Delta b],\quad (t=1,\dotsc, T-1;\ l=1,\dotsc,L),\\
\label{eqn:bct}
&\bmX_{t,1} = (x_s(t), y_s(t)) \in \widetilde{Q}_1,\quad \bmX_{t,L}= (x_g(t), y_g(t)) \in \widetilde{Q}_L, 
\end{align}
\end{itemize}
where equations \eqref{eqn:physt}, \eqref{eqn:timet}, and \eqref{eqn:bct} represent the physical constraint, time constraint, and boundary condition, respectively. $\Delta b$ represents the maximum speed of arm motion 
between $t$ and $t+1$. Small $\Delta b$ generates smooth arm motion. 
Schematics of matrix and network structure are shown in Fig.~ \ref{fig:kaiso}. 
The P and N layers at the $l^{th}$ column and $t^{th}$ row of the matrix, 
which represent the position of $l^{th}$ joint at time $t$, are denoted 
by $(t, l)^+$ and $(t, l)^-$, respectively. 

The indexes of the variables are changed as follows: 
\begin{equation*}
\begin{aligned}
&k^\pm(i,j,l,t) = (i + (j-1)J_x + (l-1)J_xJ_y + (t-1)J_xJ_yL)^\pm,\\
&\bm{\xi}_{k^\pm(i,j,l,t)} := (\xi_{k^\pm(i,j,l,t)}, \eta_{k^\pm(i,j,l,t)}),\\
&\xi_{k^\pm(i,j,l,t)} := x_{\min} + (i-1)(x_{\max}-x_{\min})/(J_x-1),\\
&\eta_{k^\pm(i,j,l,t)} := y_{\min} + (j-1)(y_{\max}-y_{\min})/(J_y-1),\\
&\Lambda^\pm(t, l) = \{(1+(l-1)J_xJ_y + (t-1)J_xJ_yL)^\pm, \cdots, (J_xJ_y+(l-1)J_xJ_y + (t-1)J_xJ_yL)^\pm\},\\
&\Lambda(t, l) = \{(1+(l-1)J_xJ_y + (t-1)J_xJ_yL), \cdots, (J_xJ_y+(l-1)J_xJ_y + (t-1)J_xJ_yL)\},\\
&\widetilde\Omega := \{\bmx \in \mathbb{R}^2\ |\ \bmx = \bm{\xi}_{k(i,j,\cdot,\cdot)}, (i,j)\in \Gamma\},
\end{aligned}
\end{equation*}
The node number at matrix $(i,j)$ in $(t, l)^+$ and $(t, l)^-$ is denoted by $k^+(i, j, l, t)$ and $k^-(i, j, l, t)$, respectively.
The feasible region restricted by the time constraint is defined as 
\begin{equation*}
\begin{aligned}
&B^+(m^+(i,j,l,t)) = \left\{k^+ \in \Lambda^+(t, l)\ |\ |\bm{\xi}_{k^+} - \bm{\xi}_{m^+}| \in [0, \Delta b], m^+\in \Lambda^+(t+1, l) \right\}, \\
&B^-(m^-(i,j,l,t)) = \left\{k^- \in \Lambda^-(t, l) \ |\ |\bm{\xi}_{k^-} - \bm{\xi}_{m^-}| \in [0, \Delta b], m^-\in \Lambda^-(t-1, l) \right\}. \\
\end{aligned}
\end{equation*}
We define respectively a set of node positions  for time $t$ that satisfy the boundary conditions and physical constraint as
$$
\widetilde{S}^\pm_{\text{B,P}}(t) \subset \Lambda^\pm(t,1)\times\cdots \times \Lambda^\pm(t,L)=:\widetilde{\Lambda}^\pm(t). 
$$
A set of plausible solutions satisfying the time constraint is defined as follows: 
\begin{align*}
&\widetilde{B}^+(\bm{k}_t^+):= B^+(k_{t,1}^+)\times \cdots \times B^+(k_{t,L}^+),\ 
\bm{k}_t^+ = (k_{t,1}^+, \dotsc, k_{t,L}^+)\in \Lambda^+(t,1)\times\cdots \times \Lambda^+(t,L). 
\end{align*}

\begin{figure}[htbp]
 \begin{center}
 \includegraphics[keepaspectratio=true, width=0.9\textwidth]{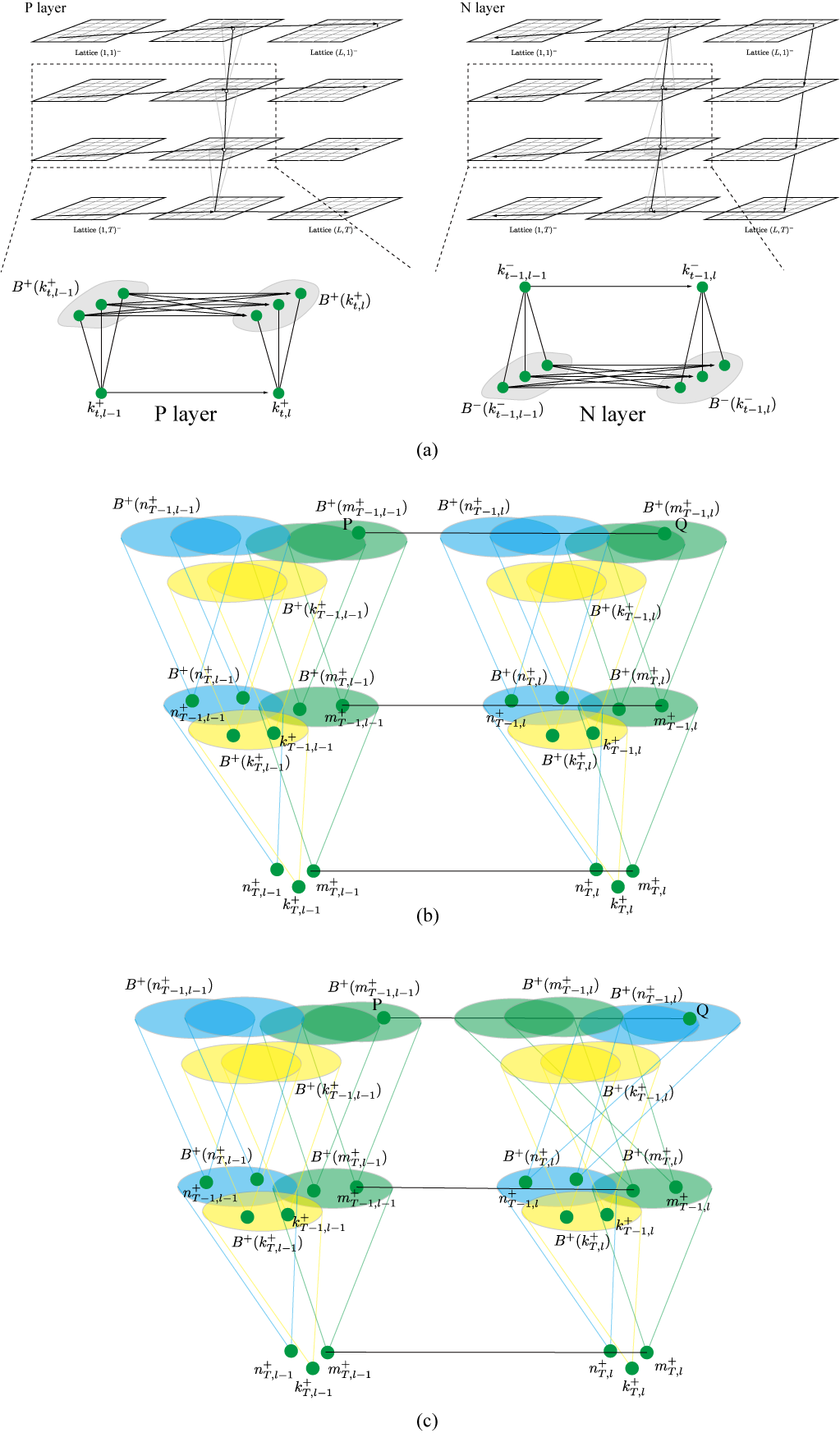}
\end{center}
 \caption{(a) Concept of the model represented by equation \eqref{eqn:modelt}. (b)
 Black lines indicate the solutions that satisfy the physical and time constraints. 
Time constraint satisfies the condition (C). In fact, point P is in $B^+(m^+_{T,l-1})$, and 
 point Q is also in $B^+(T-2; l, m^+_{T,l})$. 
 (Right) Time constraint does not satisfy the condition (C). 
In fact, point P is in $B^+(m^+_{T,l-1})$, but
 point Q is not in $B^+(m^+_{T,l})$. 
 }
\label{fig:concept-time}
\end{figure}

The network connection procedures ($\text{P2}^{\prime}$) and ($\text{P4}^{\prime}$) 
are modified as ($\text{P2}^{\prime\prime}$) and ($\text{P4}^{\prime\prime}$), 
and a procedure ($\text{P5}^{\prime\prime}$) for the time constraint 
is added. 
\begin{itemize}
\item[($\text{P2}^{\prime\prime}$)]
The links in P layer are attached if the corresponding two nodes satisfy 
the condition in equation \eqref{eqn:constd}, i.e.,
\begin{equation*}
\begin{aligned}
&a_{m^+,k^+} = 
\left\{
\begin{aligned}
&1\quad \text{if}\quad |\bm{\xi}_{m^+} - \bm{\xi}_{k^+}| \in [d_l - \Delta d_l, d_l + \Delta d_l],\ m^+\in \Lambda^+(t,l),\\
&\qquad\quad k^+\in \Lambda^+(t,l+1)\\
&0\quad \text{otherwise}
\end{aligned}
\right.
,\\
&l\in 1,\dotsc, L-1,\ t\in 1,\dotsc, T.
\end{aligned}
\end{equation*}
The connections in N layer, $a_{k^-,m^-}$ ($k^-\in \Lambda^-(t, l), m^-\in \Lambda^-(t+1, l)$) are determined according to (P2). 
\item[($\text{P4}^{\prime\prime}$)]
We denote the node number for the start (target) node in P and N layer at time $t$ as $k_s^+(t)$ ($k_g^+(t)$) and $k_s^-(t)$ ($k_g^-(t)$), respectively. 
According to ($\text{P4}^\prime$), we attach excitatory links from the target node in P layer to the target node in N layer and from the start node in N layer to the target node in P layer, i.e.,
\begin{equation*}
\hat{a}_{k_s^-(t), k_s^+(t)} = 1,\quad \hat{a}_{k_g^+(t), k_g^-(t)} = 1,\quad t=1,\dotsc, T
\end{equation*}

\item[($\text{P5}^{\prime\prime}$)]
Excitatory links are attached from node $k^+_{t,l}\in \Lambda^+(t, l)$ to $k^+_{t-1,l}\in \Lambda^+(t-1,l)$ if $k^+_{t-1,l}\in B^+(\{k^+_{t,l}\})$.  
Similarly, excitatory links are attached from node $k^-_t\in \Lambda^-(t,l)$ to $k^-_{t+1, l}\in\Lambda^-(t+1,l)$ if $k_{t+1,l}\in B^-(\{k_{t, l}\})$. 
The presence and absence of the connection is represented by
\begin{equation*}
\begin{aligned}
&b_{m^+, k^+} = 
\begin{cases}
1\qquad k^+ \in B^{+}(\{m^+\}), \\
0\qquad \text{otherwise}, 
\end{cases}\\
&b_{m^-, k^-} = 
\begin{cases}
1\qquad k^- \in B^{-}(\{m^-\}), \\
0\qquad \text{otherwise}, 
\end{cases}\\
&i=1,\dotsc, J_x; j =1,\dotsc, J_y. 
\end{aligned}
\end{equation*}
\end{itemize}
The following condition (C) is not necessary but it enables the system to find a possible solution sequentially from layer $(T, l)^\pm$ to $(1, l)^\pm$ 
($l\in [1,\dotsc, L]$). This implies that the system can quickly find a solution under the condition (C). 
\begin{itemize}
\item[(C)]
For any $t\in [2,\dotsc, T]$ and element $\bm{k}^+_{t}\in \widetilde{S}_{\text{B,P}}^+(t)$, there exists 
$\bm{k}^+_{t-1}\in \widetilde{S}^+_{\text{B,P}}(t-1)\cap \widetilde{B}^+(\bm{k}^+_{t})$ (Fig.~ \ref{fig:concept-time}(b)(c)). 
\end{itemize}

We use Model I for $t=1$ and Model II for $t=2,\dotsc, T$. 
We assume that the node $k$ receives an excitatory signal when $a_{m', k}b_{m'',k} = 1$, and both the nodes $m'$ and $m''$ 
are in the ON state, i.e., the node can be in the ON state if it satisfies the physical and the time constraints. 
The model can be expressed as follows: 
\begin{equation} \label{eqn:modelt}
\begin{aligned}
 &\dot{u}_{k^+} = f(u_{k^+},v_{k^+}) + \mu_1
 G\left(\sum_{m^{+}\in\Lambda^+} a_{m^+,k^+}
 H_0(u_{m^+}-\theta_1) \right)\times
 G\left(\sum_{m^{+}\in\Lambda^+} b_{m^+,k^+}
 H_0(u_{m^+}-\theta_1) 
 \right)\\
 &\quad + \mu_1 \hat{a}_{k^-, k^+}H_0(u_{k^-}-\theta_1) \\
 &\quad - \mu_2 G\left(\sum_{m^+\in\Lambda^+}\sum_{l\in\Lambda}
(  a_{m^+,l^+}a_{m^+,k^+}+  b_{m^+,l^+}b_{m^+,k^+})
H_0(u_{m^+}-\theta_1) H_0(u_{l^+}+u_{l^-}-\theta_2)
\right)\\
 &\quad +A\sigma_{k^+}(t),
  \vspace{0.2cm}\\
 &\dot{v}_{k^+} = g(u_{k^+},v_{k^+}), \\
 &\dot{u}_{k^-} = f(u_{k^-},v_{k^-}) + \mu_1
 G\left(\sum_{m^{-}\in\Lambda^-} a_{m^-,k^-}H_0(u_{m^-}-\theta_1)
 \right)\times
 G\left(\sum_{m^{-}\in\Lambda^-} b_{m^-,k^-}
 H_0(u_{m^-}-\theta_1) \right)\\
&\quad +\mu_1\widetilde{H}_0(u_{\hat{k}}-\theta_1; l^-)\\
 &\quad - \mu_2 G\left(\sum_{m^-\in\Lambda^-}\sum_{l\in\Lambda}
  (a_{m^-,l^-}a_{m^-,k^-}+b_{m^-,l^-}a_{m^-,k^-})
H_0(u_{m^-}-\theta_1) H_0(u_{l^-}+u_{l^+}-\theta_2)\right)\\
 &\quad +A\sigma_{k^-}(t),
  \vspace{0.2cm}\\
 &\dot{v}_{k^-} = g(u_{k^-},v_{k^-}),\\
\end{aligned}
\end{equation}%
where $k^\pm = k^\pm(i,j,l,t)\in \Lambda^\pm := \cup_{l, t} \Lambda^\pm(t,l)$ and 
\begin{equation*}
\begin{aligned}
&\widetilde{H}_0(x_{\hat{k}}; l)=
\begin{cases}
H_0(x_{\hat{k}}) \quad &(\text{if $l = L^-$})\\
0\quad &(\text{otherwise})
\end{cases}
,\\
&\hat{k} = \hat{k}(i,j,l,t) = 
\begin{cases}
k^+_g(1) \quad &(\text{if $l = L^-$ and $t=1$})\quad (\text{Model I})\\
k^-_g(t-1) \quad &(\text{if $l = L^-$ and $t\geq 2$})\quad (\text{Model II})
\end{cases}
.
\end{aligned}
\end{equation*}

The initial data is taken such that the node at the target position $\bm{\xi}_{k_g^-(1)}$ 
is in ON state. The ON state propagates according to the following sequence: 
\begin{itemize}
\item[(i)] The ON state propagates from $(1, L)^{-}$ to $(1, 1)^{-}$ layer
and from $(1, L)^{-}$ to $(T, L)^{-}$ layer (Fig.~\ref{fig:kaiso}(c)). 
\item[(ii)] The ON state propagates from $(1, l)^{-}$ layer to $(T, l)^{-}$ layer 
and from $(t, L)^{-}$ to $(t, 1)^{-}$ layer (Fig.~\ref{fig:kaiso}(c)). 
\item[(iii)] The ON state propagates from $(T, l)^{+}$ layer to $(1, l)^{+}$ layer 
and from $(t, 1)^{+}$ to $(t, L)^{+}$ layer (Fig.~\ref{fig:kaiso}(d)). 
\item[(iv)]  After the ON state reaches $(T, L)^+$,  
the nodes $\bm{k}^\pm_T\in \widetilde{S}^\pm_{\text{B,P}}(T)$ are selected during the process (iii) (Fig.~\ref{fig:kaiso}(d)). 
\item[(v)]  Due to (C), there exists $\bm{k}^\pm_{T-1}\in \widetilde{S}^\pm_{\text{B,P}}(T-1)\cap \widetilde{B}^+(\bm{k}^+_{T})$. 
\item[(vi)] The process (v) successively occurs for $t=T-2,\dotsc, 1$. 
\end{itemize}

\begin{figure}[htbp]
 \begin{center}
 \includegraphics[keepaspectratio=true, width=0.5\textwidth]{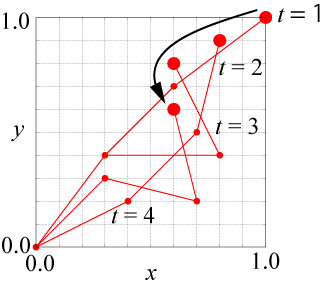}
\end{center}
 \caption{Solution of equation \eqref{eqn:modelt} displayed on $x$--$y$ plane.}
 \label{fig:timeseq}
\end{figure}

\noindent Figure \ref{fig:timeseq} shows a numerical solution of equation \eqref{eqn:modelt} 
for $L=4$ and $T=4$, where the start and target points are given by
\begin{equation*}
(x_{k^\pm_s(t)}, y_{k^\pm_s(t)}) = (0, 0)\quad (t=1,\dotsc, T)
\end{equation*}
and 
\begin{equation*}
\begin{aligned}
&(x_{k^\pm_g(1)}, y_{k^\pm_g(1)}) = (1, 1),\\
&(x_{k^\pm_g(2)}, y_{k^\pm_g(2)}) = (0.8, 0.9),\\
&(x_{k^\pm_g(3)}, y_{k^\pm_g(3)}) = (0.6, 0.8),\\
&(x_{k^\pm_g(4)}, y_{k^\pm_g(4)}) = (0.6, 0.6).
\end{aligned}
\end{equation*}
We consider that $\theta_1=1.1$, $\theta_2=3.63$, $\mu_1=1.6$, $\mu_2=9.0$, $A=1.0\times 10^{-3}$, $\Delta b=0.25$, and $J_x=J_y=11$. 

\section{Discussion} 
\begin{figure}[htbp]
 \begin{center}
 \includegraphics[keepaspectratio=true, width=0.8\textwidth]{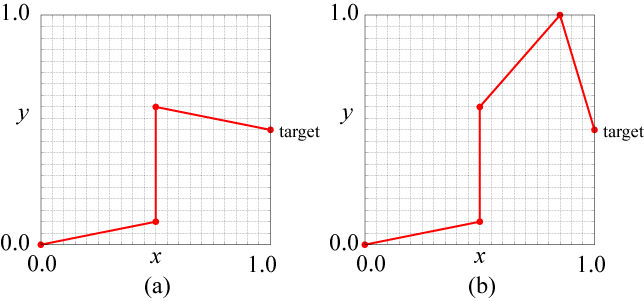}
\end{center}
 \caption{Positions of the solution of $\bm{X}_{l}$ for (a) $\tau = 400$ and (b) $\tau=1000$. 
 A new $4^{th}$ link is attached at $\tau=500$. 
 }
\label{fig:tool}
\end{figure}

Model I can be easily applied to the kinematics problem under the case that the joint number increases or decreases during computation. 
Such situations occur when a tool is being handled or an additional joint is attached. 
For example, when the joint number increases from $L$ to $L+1$, we add $L+1^{th}$ layer as well as excitatory and inhibitory links between $L^{th}$ and $L+1^{th}$ layers according to ($\text{P2}^\prime$), 
($\text{P3}$), and ($\text{P4}^\prime$). Further, we remove the existing connection at the target point in $L^{th}$ layer and attach new excitatory connection at the new target point in $L+1^{th}$ layer. In numerical simulation, we add a segment with length $l=2.0/(L-1)=0.4$ at $\tau=500$. 
$L$ is changed from $4$ to $5$, and $a_{ij}$ is set according to ($\text{P2}^\prime$) and ($\text{P3}$) during computation. 
The new target position in $L+1^{th}$ layer is $(x,y) = (1.0, 0.5)$. 
After transient dynamics, the system successfully finds a new solution (Fig.~ \ref{fig:tool}). 

A viable way to improve the accuracy of the solution is to increase $J_x$ and $J_y$. However, the number of steps increases exponentially as $J_x$ and $J_y$ increase. 
Another potential way is to successively decrease the area of the search region as the model finds a solution, but $J_x$ and $J_y$ should be constant. 
Here, we set $x_{\min}, x_{\max}, y_{\min},y_{\max}$ to $x_{\min}^l(0), x_{\max}^l(0), y_{\min}^l(0), y_{\max}^l(0)$
and replace them by $x_{\min}^l(1), x_{\max}^l(1), y_{\min}^l(1), y_{\max}^l(1)$, respectively,
such that $|x_{\max}^l(1)-x_{\min}^l(1)|$ and $|y_{\max}^l(1)-y_{\min}^l(1)|$ become smaller than $|x_{\max}^l(0)-x_{\min}^l(0)|$ and $|y_{\max}^l(0)-y_{\min}^l(0)|$ as the model finds a solution for 
a given $x_{\min}^l(0), x_{\max}^l(0), y_{\min}^l(0), y_{\max}^l(0)$. 
This algorithm is expressed as follows: 
\begin{itemize}
\item[Step 0]
Set $n\leftarrow 0$, $0<r<1$, $x_{\min}^l(n), x_{\max}^l(n), y_{\min}^l(n)$, and $y_{\max}^l(n)$. 
\item[Step 1]
Find a solution by using Model I. Let the solution be $\bmX_c = (x_c, y_c)$ and $n\leftarrow n+1$.
\item[Step 2]
$x_{\min}^l(n), x_{\max}^l(n), y_{\min}^l(n), y_{\max}^l(n)$ are given by 
\begin{equation*}
\begin{aligned}
&\Delta d_l \leftarrow r \Delta d_l, \\
&x_{\min}^l(n) = x_c - (J_x-1)\Delta d_l/2,\quad x_{\max}^l(n) = x_c + (J_x-1)\Delta d_l/2,\\
&y_{\min}^l(n) = y_c - (J_y-1)\Delta d_l/2,\quad y_{\max}^l(n) = y_c + (J_y-1)\Delta d_l/2. 
\end{aligned}
\end{equation*}
We stop the computation if $|\bmX_{l}- \bmX_{l+1}|$ become smaller than the expected precision for all $l=1,\dotsc, L-1$. 
Otherwise, we return to step 1. 
\end{itemize}
The robustness of the calculation can be enhanced if $r$ approaches $1$, but the iteration time between steps 1 and 2 increases. 
In future, we hope to derive the optimal value of $r$ for which the model robustly finds a solution at every step with minimum iteration steps. 

\begin{figure}[htbp]
 \begin{center}
 \includegraphics[keepaspectratio=true, width=0.4\textwidth]{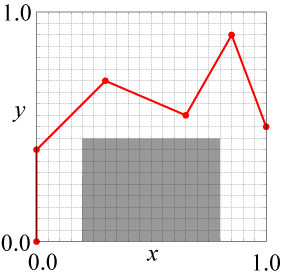}
\end{center}
 \caption{A possible solution when $a_{ij}$ is set such that each segment does not touch the forbidden region. The gray region indicates the forbidden region. 
 We consider that $\bm{X}_1= (0,0)$, $\bm{X}_L = (1,0.5)$, $L = 6$, and $l = 2.0/(L-1) = 0.4$.}
\label{fig:segmentnotouch}
\end{figure}

In our numerical experiments, we mainly considered the constraints for the position of the joints. 
We can formulate the model such that every segment does not enter the forbidden regions. 
Figure \ref{fig:segmentnotouch} shows an example of the solution of DBVP when 
the connections are determined as follows:
\begin{equation*}
\begin{aligned}
&a_{k^+,m^+} = 
\left\{
\begin{aligned}
&1\quad \text{if}\quad |\bm{\xi}_{k^+} - \bm{\xi}_{m^+}| \in [d_l - \Delta d_l, d_l + \Delta d_l], 
\text{and the line segment connecting $\bm{\xi}_{k^+}$ and $\bm{\xi}_{m^+}$ does not enter $\widetilde{Q}_l$}\\
&0\quad \text{otherwise}
\end{aligned}
\right. 
,
\end{aligned}
\end{equation*}
where $k^+\in \Lambda^+(l)$, $m^+\in \Lambda^+(l+1)$, $l=1,\dotsc, L-1$. 

\section*{Acknowledgements}
This work was supported by JSPS KAKENHI grant numbers 18H04940 and 17K05361. 
\begin{figure*}[htbp]
 \begin{center}
 \includegraphics[keepaspectratio=true, width=0.7\textwidth]{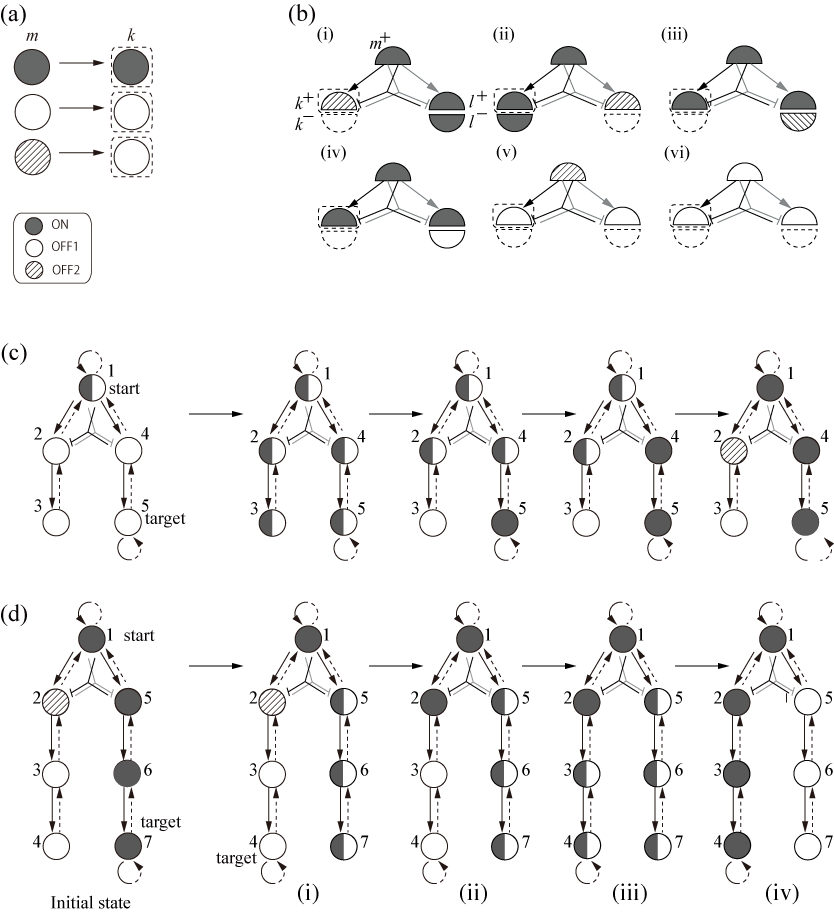}
\end{center}
 \caption{
(a) Stationary state of the node $k$ when the state of node $m$ is given. 
(b) Examples of the possible stationary states at the branching point of the excitatory links. Stationary state of the node $k^+$ when the states of nodes $m^+$, $l^+$, and $l^-$ are given. 
(c) Solution finding process of the proposed model. The arrows indicate the excitatory links. ON waves initially propagate along the excitatory links. Inhibitory interaction occurs after an ON wave in the N layer reaches the node $1^-$ (right-most panel). 
The left and right halves of the circle indicate the nodes in the P layer and the N layer, respectively. 
(d) Recovery process when the target position is changed from node $6^\pm$ to $7^\pm$. 
When the node $5^-$ acquires OFF1 state, post-inhibitory rebound occurs at the node $2^\pm$, and 
nodes $3^+$, $4^+$, $4^-$, $3^-$ and $2^-$ successively acquire ON state. 
}
\label{fig:state}
\end{figure*}
\\
\\
\\

\appendix
\section{Single node dynamics and definition of node states}\label{sec:append1}
\begin{figure}
 \begin{center}
 \includegraphics[keepaspectratio=true, width=0.9\textwidth]{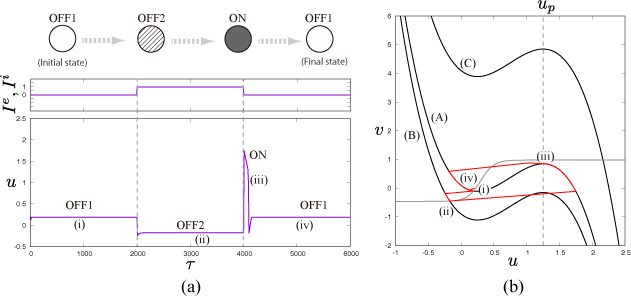}
\end{center}
\caption{
(a) Time sequence of the solution for the single-node model represented by equation \eqref{eqn:single}, where $I_e$ and $I_i$ are 
varied according to equation \eqref{eqn:ieii}. (b) The red line indicates the trajectory of the solution 
shown in (a). The solid black lines denoted by (A), (B), and (C) correspond to $u$-nullcline when $(I_e, I_i) = (1,0)$, $(0,0)$, and $(1,1)$, respectively. The gray line in (b) corresponds to $v$-nullcline, which is independent of $I_e$ and $I_i$.}
\label{fig:pir}
\end{figure}

The definition of node states and the procedure for determining the parameters in equation \eqref{eqn:model} are based on an earlier 
study 
%\cite{Ueda2015}. 
(Ueda, K. I., Yadome, M., \& and Nishiura, Y. (2015)), The dynamics of an isolated node is described by the following equation:
\begin{equation}\label{eqn:single}
\begin{aligned}
 &\dot{u} = f(u,v)+ \mu_1 I_e - \mu_2 I_i, \\
 &\dot{v} = g(u,v).
\end{aligned}
\end{equation}
where $f(u, v)=-a u^3 + b u^2 - c u + d - v$,
$g(u,v)=\epsilon [p\tanh\left((u-q)/r\right) + s - v]$.
We set $(a,b,c,d,p,q,r,s,\epsilon)=(1.92,4.32,1.8,0.1,0.72,0.3,0.2,0.261,0.03)$. 
The parameter values are set such that the system has only 
one stable stationary solution and the distance between $u$-nullcline ($f=0$) and 
$v$-nullcline ($g=0$) becomes sufficiently small
at the peak point of $u$-nullcline ($u=u_p$ in Fig.~ \ref{fig:pir}(b)). 
The terms $I_e$ and $I_i$ correspond to the excitatory and inhibitory inputs, respectively, and they are either 0 or 1. 
In equation \eqref{eqn:model}, due to the network construction, each node receives one of the following signals: $(I_e,I_i)= (0,0)$, $(1,0)$, or $(1,1)$. 
For all the cases, the model has only one stable stationary solution (Fig.~ \ref{fig:pir}(b)). 
We define the stationary state when $(I_e,I_i)\equiv (0,0)$, $(1,1)$, and $(1,0)$ 
acquire OFF1, OFF2, and ON states, respectively. The parameters $\theta_1$, $\theta_2$, $\mu_1$, and $\mu_2$ 
are set such that the system exhibits PIR and $H_0(u_{m^\pm}-\theta_1) H_0(u_{l^\pm}+u_{l^\mp}-\theta_2)=1$ when 
the nodes $m^\pm$, $l^\pm$, and $l^\mp$ are in ON state. 

\section{Pathfinding system}\label{sec:append2}
\subsection{State transition}
The fundamental role of the excitatory link is to propagate ON state and that of the inhibitory link is to select a solution at the branching point of the excitatory links. Suppose that the node $k$ receives an excitatory link from node $m$. The node $k$ acquires ON state if the node $m$ is in ON state and acquires OFF1 state for the remaining cases (Fig.~ \ref{fig:state}(a)). 
According to (P3), inhibitory links are installed at the branching point of the excitatory links. Typical cases are shown in Fig.~ \ref{fig:state}(b). It is evident that the node $k^+$ acquires OFF2 state 
when the nodes $m^+$, $l^+$, and $l^-$ are in ON state. For the remaining cases (ii)--(vi), the node $k^+$ 
does not receive inhibitory signals. 
The search and selection process is schematically shown in Fig.~ \ref{fig:state}(a). 
%For convenience, the node at position $n$ in the P and N layers is called $n^+$ and $n^-$, respectively. 
We assume that node $1^+$ is in ON state and the others are in OFF1 state. 
ON state propagates along $1^+\to 2^+ \to 3^+$ and $1^+\to 4^+ \to 5^+$, and then 
along $5^-\to 4^- \to 1^-$ in the N layer. 
Inhibition occurs when the nodes in P and N layers at point $4^\pm$ acquire ON state and that at point $2^\pm$ acquire OFF2 state. 

\subsection{Postinhibitory rebound}
PIR is crucial for the self-recovery property of the model. 
According to our network construction procedure, PIR occurs at the branching point of the excitatory links. Thus, the OFF2 state is observed when the corresponding node receives both excitatory and inhibitory signals. We show this fundamental behavior by using a simple model with external forces corresponding to excitatory and inhibitory signals. 
\begin{equation}\label{eqn:single}
\begin{aligned}
&\dot{u} = f(u,v) + \mu_1 I_e,\\
&\dot{v} = g(u,v) + \mu_2 I_i,
\end{aligned}
\end{equation}
where $I_e$ and $I_i$ are defined as follows:
\begin{equation}\label{eqn:ieii}
I_e = I_i = 
\begin{cases}
1 & \text{for}\ t \in [2000,4000],\\
0 & \text{otherwise},
\end{cases}
\end{equation}
For $\tau\in [0,2000)$, the node is in OFF1 state because it does not receive any signal. When the node receives excitatory and inhibitory signals for $\tau\in [2000, 4000]$, the solution approaches OFF2 state. When the signals are removed, the solution temporally 
approaches ON state and then finally converges to OFF1 state. 
This temporal activation is called PIR. 

\subsection{Recovery process}
Figure \ref{fig:state}(b) shows the recovery process of the model \eqref{eqn:model} when the target position is changed. After the target position is changed, nodes $7^-$, $6^-$, $5^-$, $1^-$, and $1^+$ in the N layer successively acquire OFF1 state. 
When the node $5^-$ acquires OFF1 state, the nodes $2^+$ and $2^-$ acquire ON state due to PIR, and ON state propagates along $2^+\to 3^+\to 4^+\to 4^-\to 3^-$ and $2^-\to 1^- \to 1^+$. Finally, the model finds a new path. 

\end{document}